# Thermal and thermoelectric properties of an antiferromagnetic topological insulator MnBi$_2$Te$_4$

H. Zhang,[1] C.Q. Xu,[1] S.H. Lee,[2,3] Z.Q. Mao,[2,3] and X. Ke[1]

[1] Department of Physics and Astronomy, Michigan State University, East Lansing, Michigan 48824, USA
[2] Department of Physics, Pennsylvania State University, University Park, Pennsylvania 16802, USA
[3] 2D Crystal Consortium, Materials Research Institute, Pennsylvania State University, University Park, Pennsylvania 16802, USA

The discovery of an intrinsic antiferromagnetic topological insulator (AFM-TI) in MnBi$_2$Te$_4$ has attracted intense attention, most of which lies on its electrical properties. In this paper, we report electronic, thermal, and thermoelectric transport studies of this newly found AFM-TI. The temperature and magnetic field dependence of its resistivity, thermal conductivity, and Seebeck coefficient indicate strong coupling between charge, lattice, and spin degrees of freedom in this system. Furthermore, MnBi$_2$Te$_4$ exhibits a large anomalous Nernst signal, which is associated with non-zero Berry curvature of the field-induced canted antiferromagnetic state.

Intrinsic magnetic topological materials have been intensely sought after by researchers in materials science and condense matter physics communities in the past few years [1-10]. In particular, the integration of magnetism and non-trivial band topology in magnetic topological insulators (MTIs) provides an ideal ground for identifying exotic states and quasiparticles, such as quantum anomalous Hall insulator, axion insulator and non-Abelian Majorana fermions [11-13]. Three routes to search for new MTIs have been actively explored, including doping topological insulators (TIs) with magnetic elements [14], building heterostructures by stacking TIs and magnetic layers together [15], and synthesizing single-phase materials which can host magnetism and topological bands [10]. The recent discovery of the characteristics of an intrinsic MTI in $MnBi_2Te_4$ (MBT) [8] marks a major advancement in the latter route, which further attracts avid efforts in exploring all aspects of this newly found MTI [11,13,16-21] and its derivatives [22-28].

In MBT the magnetic order is associated with the Manganese atoms, the non-trivial band topology arises from the $p_z$ bands of Bismuth and Tellurium [8]. It undergoes a paramagnetic-to-antiferromagnetic order at $T_N$ = 25 K with an A-type spin structure where spins are ferromagnetically aligned within the *ab*-plane but coupled antiferromagnetically along the *c*-axis [16,21]. And the system has a magnetic easy-axis along the *c*-direction [21]. The antiferromagnetic ground state is predicted to gap out both bulk and surface states and create an axion insulator state, while Weyl semimetal is proposed to emerge in the ferromagnetic phase [7,17]. Indeed, the observation of robust axion insulator state has been reported recently [11], and the field-induced topological phase transition has been inferred from recent electronic transport studies [29,30]. These theoretical predictions and experimental studies demonstrate that new quantum states can be achieved in a system wherein intrinsic magnetism and non-trivial band topology couples strongly.

Thus far, most studies of MBT are mainly focused on the interplay between magnetism and band topology, while the lattice degree of freedom and its potential coupling to the spin and charge degrees of freedom remain largely unexplored. In recent inelastic neutron scattering experiments, intrinsic linewidth broadening of the spin wave spectrum has been observed, which was attributed to coupling of magnons to either phonons or electrons [20]. More recently, terahertz (THZ) experiment demonstrated coherent magneto-phononic coupling in $MnBi_2Te_4$, which enables sub-picosecond control of magnetism with ultrafast laser [31]. These studies imply the strong coupling between spin and lattice degree of freedom in MBT. While recently H. H. Wang et al reported the temperature dependence of zero-field thermal conductivity and Seebeck coefficient in $Mn(Bi_{1-x}Sb_x)_2Te_4$ [32], they focused on the hole-doping effect induced by Sb substitution. In this paper, we report thermal and thermoelectric properties of MBT. We show that both thermal conductivity and Seebeck coefficient strongly depend on the applied magnetic field, indicative of the strong coupling between charge, lattice, and spin degrees of freedom in MBT. In addition, we observe a clear anomalous Nernst effect (ANE) feature, which is presumably ascribed to the non-zero Berry curvature in the field-induced canted antiferromagnetic phase.

MBT single crystals were grown using flux method [21]. Magnetic susceptibility measurements were carried out using a Superconducting Quantum Interference Device (SQUID) magnetometer, and the electronic and thermal and thermoelectric transport measurements were conducted using a Physical Property Measurement System (PPMS). Thermal and thermoelectric transport measurements were performed using a homemade sample puck designed to be compatible with the PPMS cryostat. The sample was attached to a piece of oxygen-free high conductivity copper used as the heat sink using silver epoxy, and a heater (~1 kΩ resistor) was attached to the other end of the sample. The heat current $J_Q$ was applied in the *ab* plane and the magnetic field was applied along the *c*-axis. We used three Cernox (Lakeshore Cryotronics) sensors for the temperature measurements, and the thermoelectric voltage was measured using K2182A Nanovoltmeters. The Nernst and Seebeck coefficients were obtained by anti-symmetrizing and symmetrizing the thermoelectric voltages measured in the presence of positive field and negative field, respectively.

Figure 1(a) shows the temperature dependence of zero-field cooled magnetic susceptibility measured with 1 T field applied along the crystalline c-axis ($\chi_c$) during the warming process. A sharp drop in $\chi_c$ at the Neel temperature ($T_N$ = 25 K) is clearly seen, characteristic of the onset of antiferromagnetic transition. The isothermal magnetization curves *M*(*H*) measured at various temperature from 2 K to 30 K are presented in Fig. 1(b). Below $T_N$, field induced metamagnetic transition occurs at the critical field ($H_{c1}$) which depends on the temperature. Previous transport and neutron scattering measurements showed that a canted antiferromagnetic phase emerges above $H_{c1}$ [21]. Another critical magnetic field $H_{c2}$ with larger magnitude is expected, above which MBT becomes fully polarized [21]. The nice agreement of magnetic susceptibility presented in Fig. 1 with the literature [21] affirms the good sample quality of our single crystals.

We now discus the temperature dependence of electronic, thermal, and thermoelectric properties of MBT. In Fig. 2(a) we show the Seebeck coefficient ($S_{xx}$) and resistivity ($\rho_{xx}$) of MBT measured at a field of 0 T and 9 T applied along the *c*-axis. At zero field, a kink-like transition around the Neel temperature is observed in both $S_{xx}$ and $\rho_{xx}$. A previous transport study has found that the spin fluctuation-driven scattering has a prominent effect in MBT, even well above $T_N$ [21]. Upon approaching $T_N$, spin fluctuations are expected to be stronger, leading to stronger scattering and thus a shorter mean-free-path of electrons and an increase in $\rho_{xx}$. As the long-range magnetic order develops below $T_N$, in which spins within the *ab* plane are ferromagnetically aligned, spin fluctuations and the electron-spin scattering are suppressed, resulting in kink in $\rho_{xx}$ at $T_N$ followed by a continuous decease upon decreasing the temperature. For $S_{xx}$, its negative sign over the whole measured temperature range suggests electrons as the dominant charge carriers, which is consistent with the Hall effect transport studies [21]. The negative sign of $S_{xx}$ was also previously observed, although only $S_{xx}$ above 15 K was report [32] and its magnitude is different from the values shown in Fig. 1(a), the latter of which is presumably associated with the quality of samples grown by different groups. As the temperature decreases, the entropy per electron decreases [33], leading to a decrease in $S_{xx}$; on the other hand, as will be discussed later, below $T_N$ the magnon-electron drag effect due to the coherent momentum conserving magnon-electron scattering gives rise to an increase in $S_{xx}$. As a result, the net $S_{xx}$ develops a broad bump around 14 K. At 9 T, MBT becomes a fully polarized ferromagnet, which increase the magnon lifetime and thus enhances the magnon-electron drag effect. Overall, these features of $S_{xx}$ and $\rho_{xx}$ indicate that the spin and charge degrees of freedom are intimately coupled in MBT.

Figure 2(b) shows the thermal conductivity ($\kappa_{xx}$) measured at 0 T and 9 T magnetic field ($H \parallel c$). At zero field, we observe a large increase of $\kappa_{xx}$ upon the onset of antiferromagnetic ordering at $T_N$. The total thermal conductivity in MBT has contributions from electrons, phonons, and magnons (i. e., $\kappa_{xx} = \kappa^e + \kappa^{ph} + \kappa^{mag}$). We first estimate the thermal conductivity of electrons ($\kappa^e$) using Wiedemann-Franz law ($\kappa^e = \sigma L T, L = 2.44 * 10^{-8} V^2/K^2$). The maximum electron contribution below 25 K is $0.05\ W/m\ K$, much smaller than the increase of thermal conductivity $\kappa_{xx}$ observed below $T_N$. Therefore, the main heat carriers in MBT are phonons and / or magnons. Magnons can contribute to the thermal conduction via two ways. One is that magnons serve as heat carriers, which would lead to an enhanced $\kappa_{xx}$ below $T_N$; the other is that magnons scatter phonons, giving rise to a reduction of the phonon contribution to $\kappa_{xx}$. Thus, the increase of $\kappa_{xx}$ below $T_N$ indicates the dominance of magnons as heat carriers. In addition, well below $T_N$ $\kappa_{xx}$ at 9 T is smaller than the value measured at zero field, which can be ascribed to the suppression of magnon population in the presence of magnetic field, suggesting that magnons contribute to the enhanced $\kappa_{xx}$ below $T_N$. On the other hand, the increase of $\kappa_{xx}$ near $T_N$ in the presence of 9 T is a sign of the dominant phonon scattering by magnons. As will be discussed later, the magnetic field dependence of $\kappa_{xx}$ below $T_N$ shows non-monotonic behavior, suggesting that in MBT the spin and lattice degrees of freedom are also intertwined.

The prominent effects of magnetic field on the temperature dependence of $\rho_{xx}, \kappa_{xx}, S_{xx}$ indicate that the spin, lattice, and charge degrees of freedom are strongly coupled in MBT. To better understand this coupling, we further investigate the field dependence of these three quantities. Note that a similar field dependence of $\rho_{xx}$ has been reported previously [21]. In Fig. 3(a), we plot the field dependence of $\Delta\kappa_{xx}/\kappa_{xx}(0)$ with $\Delta\kappa_{xx} = \kappa_{xx}(\mu_0 H) - \kappa_{xx}(0)$ measured at some selected temperatures. Each curve is vertically shifted for clarity. Above $T_N$, $\kappa_{xx}$ increases monotonically with magnetic field due to the suppression of spin-fluctuation driven phonon scattering. In contrast, below $T_N$, $\kappa_{xx}$ exhibits a non-monotonic field dependence. At $T = 21.5$ K, $\kappa_{xx}$ first decreases until the magnetic field reaches the first critical field $H_{c1}$ and then continues to increase with the magnetic field. As the temperature decreases, a second critical field $H_{c2}$ is observed prior to the continuous increase of $\kappa_{xx}$ with the magnetic field. Both $H_{c1}$ and $H_{c2}$ increase upon decreasing the temperature. Such a field dependence of $\kappa_{xx}$ below $T_N$ is mainly associated with the spin state of MBT in the presence of field. Previous neutron diffraction measurements show that MBT displays an A-type antiferromagnetic structure for $H < H_{c1}$ and then a canted-antiferromagnetic (CAFM) spin structure in the region of $H_{c1} < H < H_{c2}$ followed by a field-induced fully-polarized spin state above $H_{c2}$ [21]. Therefore, for $H < H_{c1}$ the application of magnetic field suppresses the magnon population, thus reducing the magnon contribution $\kappa^{mag}$ to the total $\kappa_{xx}$. For $H_{c1} < H < H_{c2}$, while the magnetic field continues to suppress the magnon population and reduces its contribution to $\kappa_{xx}$, the phonon scattering due to the magnon is reduced as well, which tends to enhance the phonon contribution to $\kappa_{xx}$. As a result of these competing effects, $\kappa_{xx}$ decreases slightly in this intermediate field region. Finally, in the fully polarized state at $H > H_{c2}$ the magnon population further decreases as the magnon gap increases with the magnetic field, which leads to the dominant phonon contribution due to the suppression of magnon-phonon interaction over the magnon contribution, thus giving rise to an enhancement in the total $\kappa_{xx}$.

Figure 3(b) shows the change of Seebeck coefficient $\Delta S_{xx}/S_{xx}(0)$ with $\Delta S_{xx} = S_{xx}(\mu_0 H) - S_{xx}(0)$ as a function of magnetic field. Above $T_N$, $S_{xx}$ increases with magnetic field; below $T_N$, $S_{xx}$ first decreases with the applied field at $H < H_{c1}$ and then sharply increase at $H_{c1}$. At $H_{c1} < H < H_{c2}$, $S_{xx}$ shows a bowl shape as a function of magnetic field; above $H_{c2}$, $S_{xx}$ continues to increase with field again. Can the variation in $S_{xx}$ in the presence of magnetic field arise from the magnetic field effect of the electron diffusion contribution? In metallic systems, one may anticipate that diffusion thermoelectric effect follows Mott relation, $S_{xx} = -\frac{\pi^2 k_B^2 T}{3 e \sigma_{xx}} \frac{d\sigma_{xx}}{d\zeta}\big|_\mu$ ($\mu$ is the Fermi energy). Assuming that the energy derivative of conductivity at Fermi energy $\frac{d\sigma_{xx}}{d\zeta}\big|_\mu$ is constant, one would expect $S_{xx}$ to increase with the increase of resistivity $\rho_{xx}$ ($= 1/\sigma_{xx}$), which is opposite to the trends of field dependence of $S_{xx}$ and $\rho_{xx}$ shown in Fig. 3(b,c). Thus, one potential scenario is that the $\frac{d\sigma_{xx}}{d\zeta}\big|_\mu$ term is field-dependent, having an opposite trend to and dominating over the field dependence of $\rho$ such that $S_{xx}$ increases while $\rho_{xx}$ decreases with field. Alternatively, the magnetic field dependence of $\Delta S_{xx}/S_{xx}(0)$ shown in Fig. 3(b) may be associated with the magnon-electron drag effect. In the presence of a temperature gradient, magnons may scatter electrons, causing a drag in electron velocity and enhancing $S_{xx}$. The initial decrease of $S_{xx}$ at $H < H_{c1}$ is likely due to the reduction of magnon population and the resulting suppression of the magnon-electron drag process. Across $H_{c1}$, the magnetic structure changes from an A-type antiferromagnet to a canted antiferromagnet [21]. As a result, the magnon relaxation lifetime $\tau_m$ and the lifetime of electron scattered by magnon ($\tau_{em}$) are altered. In the leading order approximation, the magnon-electron drag thermoelectric effect is proportional to $\tau_m/\tau_{em}$ [34]. Thus, a potential dominant increase of $\tau_m$ compared to that of $\tau_{em}$ in the canted antiferromagnet state can lead to an enhanced $S_{xx}$. At $H_{c2}$ MBT becomes a fully polarized ferromagnet [21], which can reduce the magnon scattering at the magnetic domain walls and consequently further increases $\tau_m$ and magnon-electron drag effect in $S_{xx}$. Another possible mechanism that might account for the field dependence of $S_{xx}$ is associated with the field-induced electronic structure change in MBT. That is, MBT undergoes topological phase transition from antiferromagnetic topological insulator to Weyl semimetal in the ferromagnetic state [7,17,30]. Further theoretical investigation is desirable to pin down the underlying physics of the observed $S_{xx}(H)$ behavior.

Finally, we discuss the transverse thermoelectric transport, i.e., anomalous Nernst effect (ANE) in MBT. Recently, there has been increasing interest in the ANE effect in topological materials [35-39]. ANE is complementary to anomalous Hall effect (AHE) measurements in characterizing the Berry curvature distribution in the reciprocal space [38]. Recent AHE studies suggested that a non-zero Berry curvature is created in the canted AFM state, leading to an intrinsic AHE [21]. In Fig. 4 we present the magnetic field dependence of Nernst coefficient ($-S_{xy}$) measured at various temperatures. The inset illustrates the schematic experimental setup and the sign convention here follows a previous ANE study on a Dirac semimetal $Cd_3As_2$ [39]. Overall, the Nernst signal tracks well with the magnetization data [Fig. 1(b)] in behavior. Below $T_N$, the Nernst coefficient slowly increases with the magnetic field until $H_{c1}$ at which a sudden increase in Nernst coefficient is observed, consistent with the emergence of non-zero Berry curvature in the momentum space in the canted antiferromagnetic phase suggested previously based on AHE measurements [21]. As the magnetic field is increased, MBT becomes fully polarized at $H_{c2}$ [21] and turns into a Weyl semimetal [7,17,29,30]. Nevertheless, recent electronic transport studies showed that the Weyl nodes in the polarized ferromagnetic state are not close to the Fermi energy in the pristine compound, in contrast to the lightly hole-doped Mn(Bi$_{1-x}$Sb$_x$)$_2$O$_4$ sample [29]. As a result, the slight further enhancement in $|S_{xy}|$ above $H_{c2}$ is mainly ascribable to the normal Nernst contribution instead of field-induced electronic structure change.

In summary, we have studied electronic, thermal and thermoelectric transport properties of MnBi$_2$Te$_4$. We observe both temperature and magnetic field dependence of longitudinal resistivity, thermal conductivity, and Seebeck coefficient, indicating strong coupling among charge, lattice, and spin degrees of freedom in this system. Furthermore, the observation of anomalous Nernst effect implies the creation of Berry curvature in the momentum space in the field-induced canted antiferromagnetic state.

Acknowledgement

H.Z. and X.K. acknowledge the financial support by the U.S. Department of Energy, Office of Science, Office of Basic Energy Sciences, Materials Sciences and Engineering Division under DE-SC0019259. C.X. is partially supported by the Start-up funds at Michigan State University. The financial support for sample preparation was provided by the National Science Foundation through

Figure 1

H. Zhang et al,

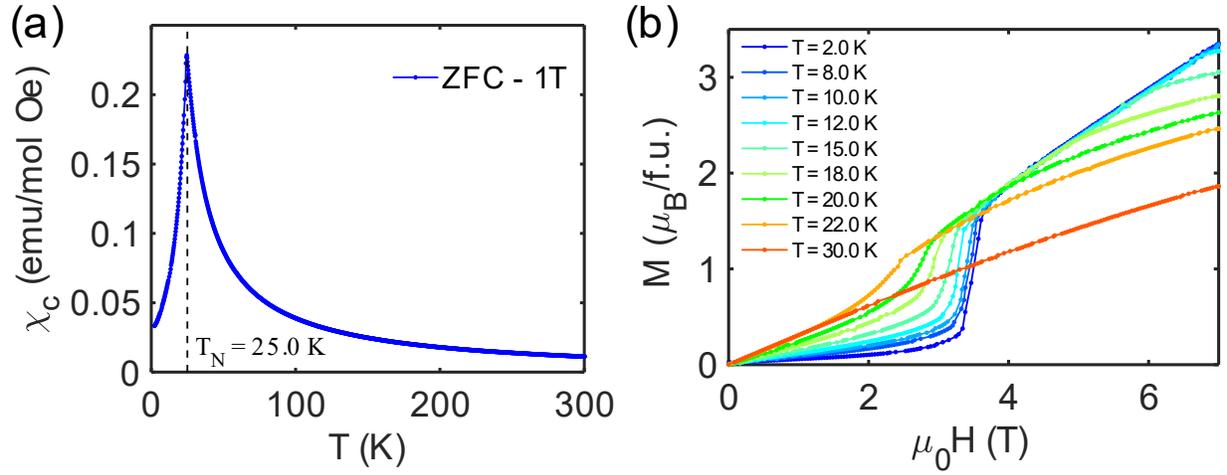

**Figure 1.** (a) Temperature dependence of magnetic susceptibility of MnBi$_2$Te$_4$ measured with a 1T magnetic field applied along the crystalline *c*-axis. The sample is cooled under zero-field conditions. (b) Isothermal magnetization measurement of MnBi$_2$Te$_4$ at various temperatures from 2 K to 30 K.

Figure 2

H. Zhang et al,

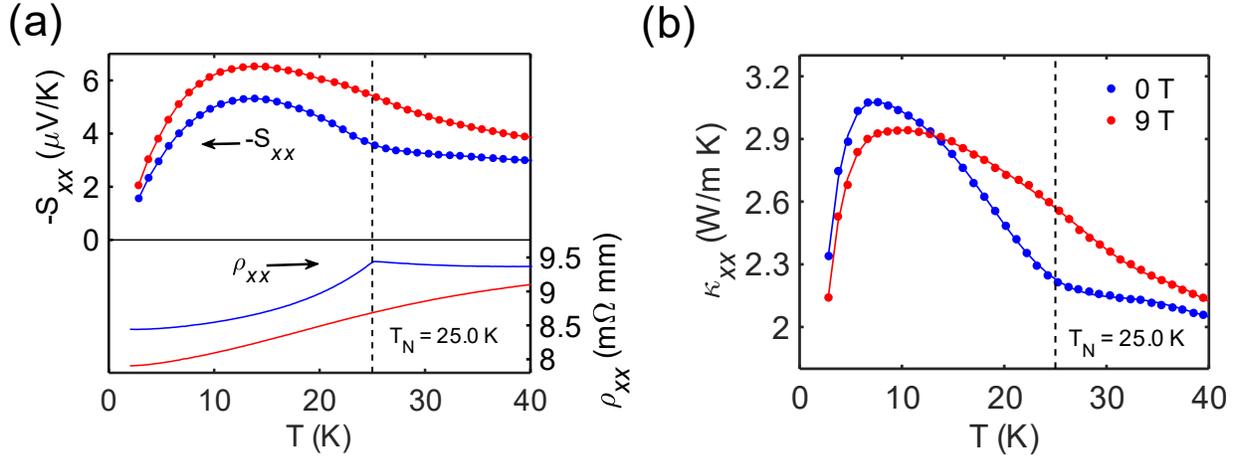

**Figure 2.** (a) Temperature dependence of longitudinal Seebeck coefficient $S_{xx}$ (upper panel) and resistivity $\rho_{xx}$ (lower panel) of MnBi$_2$Te$_4$ measured under an applied field of 0 T (in blue) and 9 T (in red, $H//c$). (b) Temperature dependence of longitudinal thermal conductivity $\kappa_{xx}$ measured under an applied field of 0 T (in blue) and 9 T (in red, H//c).

Figure 3

H. Zhang et al,

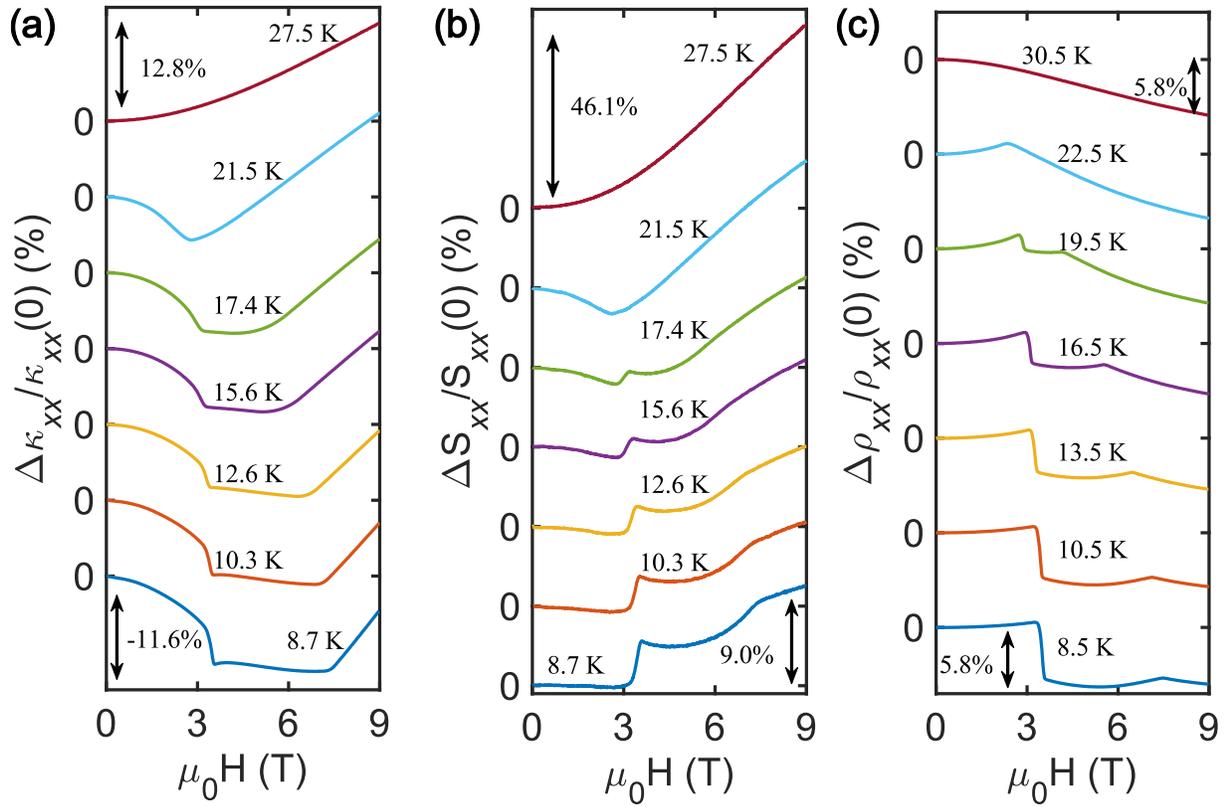

**Figure 3.** The magnetic field dependence of relative change in longitudinal (a) thermal conductivity $\kappa_{xx}$, (b) Seebeck coefficients $S_{xx}$, and (c) resistivity $\rho_{xx}$ measured at various temperatures.

Figure 4

H. Zhang et al,

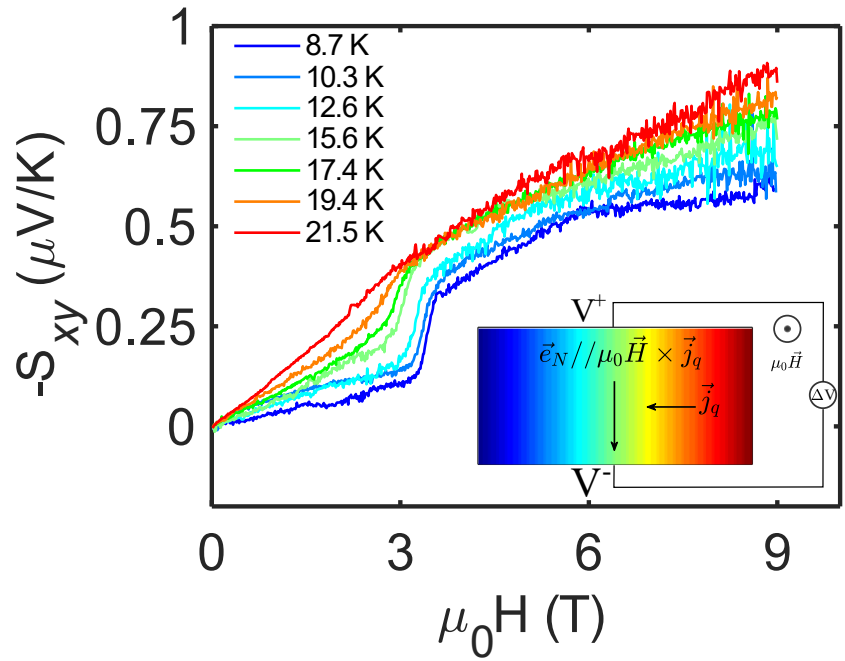

**Figure 4.** Magnetic field dependence of Nernst coefficients of MnBi$_2$Te$_4$ measured at various temperatures. The inset shows a schematic of the experimental set-up.